\documentclass[aps,prl,reprint,groupedaddress,showkeys]{revtex4-2}

\usepackage[utf8]{inputenc}

\usepackage[T1]{fontenc}
\usepackage{verbatim}
\usepackage{amssymb,amsmath}
\usepackage{bbm}
\usepackage{comment,xcolor}
\usepackage[colorlinks,linktocpage,linkcolor=blue,citecolor=red,urlcolor=red]{hyperref}
\usepackage{graphicx,float}
\usepackage{tikz}
\usepackage{appendix}
\usepackage{amsthm}
\usepackage{amssymb}
\usepackage{physics}
\usepackage{mleftright}

\usepackage{tikz}
\usetikzlibrary{3d,calc,decorations.pathreplacing}
\usepackage{tikz-3dplot}

\mleftright

\begingroup
\global\mathcode`/=\numexpr "4000 + \count2\relax
\endgroup

\def\be{\begin{equation}}
\def\ee{\end{equation}}

\renewcommand{\abs}[1]{\mathopen|#1\mathclose|}

\let\oldsection\section
\renewcommand{\section}[1]{\oldsection{\texorpdfstring{\NoCaseChange{#1}}{#1}}}

\begin{document}

\title{The IKKT renormalization group flow is IIB: \\
toward zero-d holography}

\author{Xiangwen Guan$^a$}
\author{Joel Karlsson$^{b,c}$}
\author{S{\'e}bastien Reymond$^b$}
\author{Thomas Van Riet$^b$}

\affiliation{%
    \vspace*{2ex}
    $^a$The Oskar Klein Centre and Department of Physics, Stockholm University, AlbaNova, 106 91 Stockholm, Sweden\\
    $^b$Institute for Theoretical Physics, KU Leuven, Celestijnenlaan 200D, B-3001 Leuven, Belgium\\
    $^c$Leuven Gravity Institute, KU Leuven, Celestijnenlaan 200D box 2415, B-3001 Leuven, Belgium
}

\begin{abstract}
Supergravity solutions describing stacks of D$p$-branes with $p\neq 3$ feature a non-constant dilaton profile, which is holographically mapped to the running of the SYM coupling in $(p+1)$ dimensions. For D-instantons ($p=-1$), the lack of space and time in the IKKT matrix model makes such an interpretation difficult at first. In this letter, we propose a method to achieve this based on two closely related concepts: the IKKT method of integrating out heavy strings in a Coulomb branch vacuum and the matrix RG flow of Br\'ezin and Zinn-Justin (BZJ). The notable difference between the two is that the BZJ RG flow also integrates over the Coulomb branch position. We first apply the Coulomb branch method and by relating the coefficient of the leading correction to the IKKT action with the string coupling, we can compute its dependence on the Coulomb branch position, finding a match between matrix theory and supergravity. Next, we apply the BZJ-flow to the IKKT partition function, which leads to a running of the coupling constant $g$ with the rank $N$ of the matrix model, reproducing the $N$-dependence of the axio-dilaton field in supergravity.
\end{abstract}

\maketitle

\section{Introduction}
In holography the RG scale of a field theory in $D$ spacetime dimensions is geometrized as an extra dimension in a bulk spacetime with dynamical gravity. For a conformal field theory on Minkowski spacetime this extra dimension combines into a $D+1$-dimensional Anti-de Sitter (AdS) space.
However, in the well-established holographic dualities, more space dimensions emerge in a compact space $X$ such that the dimension of AdS$_{D+1}\times X$ is 10 or 11, as required for an embedding in critical string- or M-theory.

When holography is extended to brane worldvolume theories that are non-conformal (D$p$-branes with $p\neq 3$), the spacetime profile of the string coupling is tied to the RG flow of the field theory.  The goal of this paper is to address whether and how this works for $p=-1$, i.e.\ for the D-instantons of IIB string theory whose worldvolume is described by the Euclidean IKKT matrix model \cite{Ishibashi:1996xs}
\begin{align}
 Z &= \mathcal{N}(N,g)
 \int \prod_{m=1}^{10} \dd X^{m} \prod_{\alpha=1}^{16} \dd \psi_{\alpha}\, e^{-S_{\mathrm{IKKT}}}\, ,\nonumber\\
 S_\mathrm{IKKT} & = - \frac{1}{4g^2} \Tr [X, X]^2 - \frac{1}{2g^2} \Tr \bigl(\psi \Gamma_m [X^m, \psi]\bigr)\,,
\end{align}
with $\mathcal{N}(N,g)$ a normalization factor fixed by D-instanton physics and whose details are discussed below (see \footnote{We adopt the notation $[X, X]^2 = [X_m, X_n] [X^m, X^n]$. The Weyl spinors $\psi$ have $16$ complex components and $\psi \Gamma_m [X^m, \psi] =  \psi_\alpha (\mathcal{C} \Gamma^m)^{\alpha \beta} [X^m,\psi_\beta]$. The convention for the fermion term is discussed in \cite{Hartnoll:2024csr,Komatsu:2024bop}.} for notation).
If holography applies to all D$p$-branes then we expect this matrix model to be holographically dual to IIB string theory in a background corresponding to the geometry of the same stack \cite{Gibbons:1995vg}. Note that IKKT already conjectured the matrix model to be dual to IIB, but then in flat space and in a non-holographic manner. The link between the IKKT matrix model and emergent spacetime is a topic with a rich history to which we can do no justice here and instead refer the reader to \cite{Nishimura:2020blu, Brahma:2022dsd, Steinacker:2024unq,Steinacker:2026jzp} and references therein.

The geometry of a stack of D-instantons is described using a spherical coordinate system in 10d with a radial direction $r$ and angles forming a 9-sphere $S^9$. It was recently shown that IKKT can reproduce the sphere angles holographically \cite{Ciceri:2025maa, Ciceri:2025wpb, Ciceri:2026hxr}. We aim at understanding how the radial coordinate emerges.
Since matrix models like IKKT have no space or time, explaining the radial coordinate through the notion of RG flow seems impossible. However, one intuitive way to RG flow matrix models is to reduce $N$ (the rank of the matrices), leading to an effective reduction of degrees of freedom. This idea is due to Br\'ezin and Zinn-Justin \cite{Brezin_ZJ_1992} and was developed in follow-up papers \cite{Higuchi:1993pu, Zinn-Justin:2014wva, Ydri:2020efr, Lahoche:2019ocf,Narayan:2002gv}.
We propose two ways of reducing $N$ in IKKT. First, we go on the Coulomb branch, splitting the D-instanton stack in two, and integrate out one of the stacks. The leading correction in the effective action on the remaining stack can then be used to extract the dilaton profile in the supergravity solution. Second, we apply the matrix RG flow method proposed in \cite{Brezin_ZJ_1992} to IKKT, and compute the $N$-flow of the tension term. This leads to an $N$-flow of the IKKT coupling constant and $\theta$ angle matching the $N$-dependence of the string coupling and RR axion in the bulk.

\section{Gauge couplings in holography}
We first recall some basic aspects of non-conformal holography \cite{Itzhaki:1998dd, Boonstra:1998mp, Peet:1998wn, Kanitscheider:2008kd, Bobev:2025idz}. Backreacting stacks of D$p$-branes have a string frame metric and dilaton given by
\begin{align}
&ds^2 = \frac{1}{\sqrt H} dx_{p+1}^2 + \sqrt{H}[dr^2 + r^2d\Omega_{8-p}^2]\nonumber\,,\\
&e^{\phi} = e^{\phi_0}H^{\frac{3-p}{4}}\,,
\end{align}
where $x^\mu$, $\mu = 0, \dots, p$, are the worldvolume coordinates, $H$ is the harmonic function on the space orthogonal to the brane and $\phi = \phi_0$ at spatial infinity. The spherical symmetric harmonic reads ($p<7$)
\begin{equation} \label{eq:Dp_harmonic}
H = c + \frac{Q_N}{r^{7-p}}\,,
\end{equation}
with $Q_N$ proportional to the number of branes $N$. The number $c$, when non-zero, can be fixed to $1$ by a coordinate choice. With this choice, $e^{\phi_0}$ is the string coupling at infinity, henceforth denoted by $g_s$, and
\begin{equation}
    Q_N = q_p g_s N (\alpha')^{\frac{7-p}{2}}\,,\quad
    q_p = (2\sqrt{\pi})^{5-p}\, \Gamma\left(\frac{7-p}{2}\right)\,.
\end{equation}
Taking the decoupling limit by sending $\alpha' \to 0$ with $r/\alpha'$ fixed, $c$ is negligible compared to the $r$-dependent term, and we obtain the near-horizon solution, which, by itself, does not asymptote to flat space.
We regard this as the holographic background since the massive open string states do not survive (with an important exception discussed below).
For these near-horizon solutions, we have
\begin{equation}\label{dilaton}
e^{\phi} = g_s Q_N^{\frac{3-p}{4}}r^{-\frac{(3-p)(7-p)}{4}} .
\end{equation}
On the field theory side, the effective dimensionless coupling of SYM at energy scale $E$ is
\begin{align}
    g_{\text{eff}}^2 = g_{\text{YM}}^2 N E^{p-3}
\end{align}
Following \cite{Peet:1998wn}, one can see that a supergravity mode in the bulk at radius $r$ corresponds to a QFT energy
\begin{align}
    E =\frac{(r/\alpha')^{\tfrac{5-p}{2}}}{g_{\text{YM}}N^{1/2}}\, .
\end{align}
Here we see the common intuition that RG flow in the QFT is geometrized as the radial direction in the bulk. One can obtain this relation by rescaling $r\to \gamma r$ and $x^\mu \to \gamma^{(p-5)/2} x^\mu$ so that the metric is rescaled by an overall factor $\gamma^{(p-3)/2}$. Since the field theory energy scales inversely to the worldvolume coordinates, $E \to \gamma^{(5-p)/2} E$,  this fixes $E \propto r^{(5-p)/2}$.
This is related to the scaling similarity of the supergravity recently discussed in \cite{Biggs:2023sqw,Bobev:2025idz}.
Moreover, combining the two equations above one gets
\begin{align}
    g_{\text{eff}}^2 = \left[g_{\text{YM}}N^{1/2} \left(\frac{r}{\alpha'}\right)^{\tfrac{p-3}{2}}\right]^{5-p}\, .
\end{align}
The fact that SYM in dimension $D<4$ is strongly coupled in the IR is seen here by going to small $r$, and similarly the UV is weakly coupled, since $g_{\text{eff}}$ is small at large $r$. The critical case $p=3$ is consistent with SYM in 4D being a CFT with no RG running.

Clearly one cannot just extrapolate the above equations to $D=0$ ($p=-1$) since the worldvolume becomes a point and the theory a random matrix theory; in the absence of space and time there is no standard notion of RG flow. However, one can find a notion of radial distance by going on the Coulomb branch on the matrix model side.

\section{Radial profile from the Coulomb branch}

In what follows we induce a change of rank by picking a vacuum on the Coulomb branch and integrating out the massive vector fields. From a spacetime viewpoint, it means we integrate out long strings that stretch between the stacks, as depicted in the figure below. Crucially, there is no integral over the Coulomb branch position in the path integral---in the context of QFT, this is because the VEV is a fixed boundary condition, while in IKKT this has to be implemented by hand.
\begin{figure}[H]
    \centering
    \pgfmathsetmacro{\TikzStackScale}{0.65}
    \begingroup

\providecommand{\TikzStackScale}{1}

\tdplotsetmaincoords{80}{130}

\definecolor{myyellow}{RGB}{210, 175, 55}
\definecolor{myblue}{RGB}{31, 171, 231}
\colorlet{braneColor}{myblue!60!white}
\colorlet{braneEdge}{myblue!70!black}

\begin{tikzpicture}[tdplot_main_coords,
                    scale=\TikzStackScale,
                    line cap=round,
                    line join=round]


\def\N{4}   
\def\M{5}   

\def\StackSep{10}      
\def\StackDepth{0.22}  

\def\BraneH{5.0}
\def\BraneW{3.2}

\pgfmathsetmacro{\MFx}{(\M-1)*\StackDepth/2}   


\newcommand{\Brane}[3]{%
\filldraw[
    fill=braneColor,
    draw=braneEdge,
    line width=0.5pt
]
(#1,#2,#3)
--
(#1,#2+\BraneW,#3)
--
(#1,#2+\BraneW,#3+\BraneH)
--
(#1,#2,#3+\BraneH)
-- cycle;
}

%

\newcommand{\Handle}{3.0}   
\newcommand{\Wig}{0.9}      

\newcommand{\Cubic}[3]{
\draw[#1] (#2) .. controls ($(#2)+(\Handle,0,\Wig)$) and ($(#3)+(-\Handle,0,-\Wig)$) .. (#3);
}
\newcommand{\CubicB}[3]{
\draw[#1] (#2) .. controls ($(#2)+(\Handle,0,-\Wig)$) and ($(#3)+(-\Handle,0,\Wig)$) .. (#3);
}
\newcommand{\CubicFrom}[4]{
\coordinate (cp1) at ($(#2)+(\Handle,0,\Wig)$);
\coordinate (cp2) at ($(#3)+(-\Handle,0,-\Wig)$);
\coordinate (cqa) at ($(#2)!#4!(cp1)$);
\coordinate (cqb) at ($(cp1)!#4!(cp2)$);
\coordinate (cqc) at ($(cp2)!#4!(#3)$);
\coordinate (cqd) at ($(cqa)!#4!(cqb)$);
\coordinate (cqe) at ($(cqb)!#4!(cqc)$);
\coordinate (cqf) at ($(cqd)!#4!(cqe)$);
\draw[#1] (cqf) .. controls (cqe) and (cqc) .. (#3);
}
\newcommand{\CubicFromB}[4]{
\coordinate (cp1) at ($(#2)+(\Handle,0,-\Wig)$);
\coordinate (cp2) at ($(#3)+(-\Handle,0,\Wig)$);
\coordinate (cqa) at ($(#2)!#4!(cp1)$);
\coordinate (cqb) at ($(cp1)!#4!(cp2)$);
\coordinate (cqc) at ($(cp2)!#4!(#3)$);
\coordinate (cqd) at ($(cqa)!#4!(cqb)$);
\coordinate (cqe) at ($(cqb)!#4!(cqc)$);
\coordinate (cqf) at ($(cqd)!#4!(cqe)$);
\draw[#1] (cqf) .. controls (cqe) and (cqc) .. (#3);
}
\newcommand{\CubicTo}[4]{
\coordinate (cp1) at ($(#2)+(\Handle,0,\Wig)$);
\coordinate (cp2) at ($(#3)+(-\Handle,0,-\Wig)$);
\coordinate (cqa) at ($(#2)!#4!(cp1)$);
\coordinate (cqb) at ($(cp1)!#4!(cp2)$);
\coordinate (cqc) at ($(cp2)!#4!(#3)$);
\coordinate (cqd) at ($(cqa)!#4!(cqb)$);
\coordinate (cqe) at ($(cqb)!#4!(cqc)$);
\coordinate (cqf) at ($(cqd)!#4!(cqe)$);
\draw[#1] (#2) .. controls (cqa) and (cqd) .. (cqf);
}
\newcommand{\CubicToB}[4]{
\coordinate (cp1) at ($(#2)+(\Handle,0,-\Wig)$);
\coordinate (cp2) at ($(#3)+(-\Handle,0,\Wig)$);
\coordinate (cqa) at ($(#2)!#4!(cp1)$);
\coordinate (cqb) at ($(cp1)!#4!(cp2)$);
\coordinate (cqc) at ($(cp2)!#4!(#3)$);
\coordinate (cqd) at ($(cqa)!#4!(cqb)$);
\coordinate (cqe) at ($(cqb)!#4!(cqc)$);
\coordinate (cqf) at ($(cqd)!#4!(cqe)$);
\draw[#1] (#2) .. controls (cqa) and (cqd) .. (cqf);
}


\coordinate (L1) at ( 0.44,2.4,4.00);  
\coordinate (L2) at ( 0.00,2.0,3.00);
\coordinate (L3) at (-0.44,1.2,1.35);  
\coordinate (L4) at (-0.22,1.8,0.80);

\coordinate (R1) at (10.11,1.0,3.7);
\coordinate (R2) at ( 9.89,1.5,2.9);
\coordinate (R3) at ( 9.67,0.9,2.0);
\coordinate (R4) at (10.00,0.6,1.0);

\coordinate (Ltop) at (0,\BraneW/2,\BraneH);
\coordinate (Rtop) at (\StackSep,\BraneW/2,\BraneH);


\pgfmathtruncatemacro{\Mm}{\M-1}
\foreach \i in {0,...,\Mm}{
  \pgfmathsetmacro{\x}{-(\M-1)*\StackDepth/2+\i*\StackDepth}
  \Brane{\x}{0}{0}
}


\Cubic{line width=0.6pt}{L1}{R1}                
\CubicFrom{line width=0.6pt}{L2}{R3}{0.049}     
\CubicFromB{line width=0.6pt}{L3}{R2}{0.098}
\CubicFrom{line width=0.6pt}{L4}{R4}{0.073}


\pgfmathtruncatemacro{\Nm}{\N-1}
\foreach \i in {0,...,\Nm}{
  \pgfmathsetmacro{\x}{\StackSep-(\N-1)*\StackDepth/2+\i*\StackDepth}
  \Brane{\x}{0}{0}
}


\CubicTo{line width=0.6pt,dashed}{L2}{R3}{0.049}
\CubicToB{line width=0.6pt,dashed}{L3}{R2}{0.098}
\CubicTo{line width=0.6pt,dashed}{L4}{R4}{0.073}
\CubicFrom{line width=0.6pt,dashed}{L1}{R1}{0.72}
\CubicFrom{line width=0.6pt,dashed}{L2}{R3}{0.745}
\CubicFromB{line width=0.6pt,dashed}{L3}{R2}{0.78}
\CubicFrom{line width=0.6pt,dashed}{L4}{R4}{0.698}

\foreach \p in {L1,L2,L3,L4,R1,R2,R3,R4}{\fill (\p) circle (1.2pt);}


\draw[latex-latex] (Ltop) -- (Rtop) node[midway,above] {$\Delta$};

\def\bracedrop{.1}                              
\pgfmathsetmacro{\Nh}{(\N-1)*\StackDepth/2}     
\pgfmathsetmacro{\bracepad}{.05}

\draw[decorate,decoration={brace,mirror,amplitude=4pt}]
(\MFx+\bracepad,\BraneW,-\bracedrop) -- (-\MFx-\bracepad,\BraneW,-\bracedrop)
node[midway,below=3pt] {$M$};

\draw[decorate,decoration={brace,mirror,amplitude=4pt}]
(\StackSep+\Nh+\bracepad,\BraneW,-\bracedrop) -- (\StackSep-\Nh-\bracepad,\BraneW,-\bracedrop)
node[midway,below=3pt] {$N$};

\end{tikzpicture}

\endgroup%
    \caption{An illustration of the two stacks of D$p$-branes separated by a distance $\Delta$.}
    \label{fig:stack}
\end{figure}
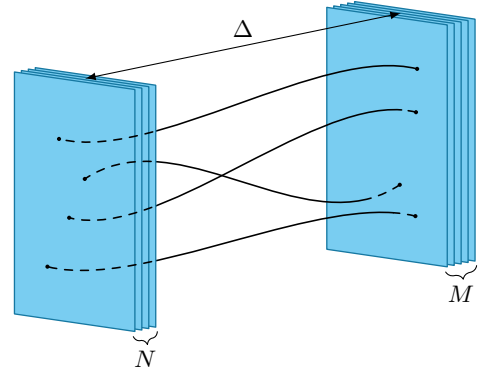

In the context of D3-brane holography this is for instance investigated in \cite{Costa:2000gk} (and references therein). We will closely follow \cite{Costa:2000gk} where the author considers two stacks of D3-branes, with respectively $N$ and $M$ branes (Fig.~\ref{fig:stack}).
The $p$-brane near-horizon two-centered solution is specified by the harmonic
\begin{equation} \label{eq:two-centered_H}
    H = \frac{Q_M}{\abs{\vec{r} - \vec{\Delta}/2}^{7-p}} + \frac{Q_N}{\abs{\vec{r} + \vec{\Delta}/2}^{7-p}}\,,
\end{equation}
where $\vec{\Delta}$ is the separation between the two stacks. We keep $\Delta/\alpha'$ fixed when taking the decoupling limit, meaning that the strings stretching between the stacks survive with finite mass. Specializing now to $p=3$, and further zooming in on the geometry near the stack of $N$ branes, the spacetime solution is approximated by the usual D3-brane geometry but with a harmonic given by
\begin{equation} \label{eq:deformed_H}
H \approx c + \frac{Q_N}{r^4}\,,\quad  c= \frac{Q_M}{\Delta^4}\,,
\end{equation}
after shifting the coordinates $\vec{r} \to \vec{r} - \vec{\Delta}/2$.
Clearly the $c$-parameter is a violent deformation of the asymptotic AdS geometry and it is conjectured to correspond to an irrelevant deformation of $\mathcal{N}=4$ SYM with a dimension 8 operator $\mathcal{O}_8 = \Tr F^4$ (index structure in \footnote{$\Tr F^4 = \frac{1}{3} \Tr(F_{mn} F^{np} F_{pq} F^{qm}) + \frac{2}{3} \Tr(F_{mn} F_{pq} F^{mq} F^{pn}) - \frac{1}{6} \Tr(F_{mn} F^{mn} F_{pq} F^{pq}) - \frac{1}{12} \Tr(F_{mn} F_{pq} F^{mn} F^{pq})$}):
\begin{equation}\label{Costa-deformation}
    \mathcal{L} = \mathcal{L}_{\text{SYM}} + \tilde{c} \Tr F^4\,.
\end{equation}
This deformation is obtained from $\mathcal{N}=4$ SYM in the following way. Starting from a $SU(N+M)$ gauge group, one goes on the Coulomb branch by giving the scalar fields a VEV $\langle X \rangle\propto\Delta$ that breaks $SU(N+M)\to SU(N)\times SU(M)$. The heavy W-bosons can be integrated out and, at one-loop level, the leading deformation to the effective action is the irrelevant operator $\Tr F^4$, which has the same form for both the $SU(N)$ and $SU(M)$ sectors. There is also a double trace operator $\Tr_N F^2 \Tr_M F^2$, but its coupling is suppressed by a factor of $M^{-1}$ with respect to $\tilde c$ in the asymptotic limit $M\to \infty$.

The expectation is that the higher-derivative operators in the non-Abelian DBI action of a stack of $N$ D-branes can be recovered this way and this leads to an argument for the holographic dictionary that relates the $c$ in the harmonic to the coupling of $\mathcal{O}_8$: when expanding the DBI action for the stack of $N$ branes at a distance $\Delta$ away from the backreacting stack of $M$ branes, one arrives at the same operator and coefficient \cite{Costa:2000gk}
\begin{equation}
   \tilde c = - \frac{(2\pi \alpha')^4 T_3 }{8 g_s}c = - \frac{\pi^2 M}{(\Delta/\alpha')^4}\, ,
\end{equation}
where we used the $p$-brane tension defined as $T_p = (2\pi)^{-p} (\alpha')^{-\frac{p+1}{2}}$.
As expected, the irrelevant deformation is suppressed in the IR ($\Delta/\alpha' \to \infty$).
In a later section, we will demonstrate how for all D$p$-branes this operator is expected to couple to the harmonic $H$ defining the string frame metric.

We now apply the same logic to IKKT holography. We go on the Coulomb branch by separating two stacks of $N$ and $M$ D-instantons, respectively, by a distance $\Delta$ and then integrate out the heavy strings. The local spacetime solution close to the stack of $N$ branes is now given by the deformed harmonic (cf.\ \eqref{eq:deformed_H})
\begin{equation}\label{D-1harm}
    H = c + \frac{Q_N}{r^8}\,,
    \quad c = \frac{Q_M}{\Delta^8}\,.
\end{equation}
In matrix language it means we integrate out rows and columns and do not integrate over the relative position $\Delta_{\mu}$ of the stacks (defining the Coulomb branch position). This gives us a \emph{deformed IKKT model}. We then compute the leading deformation operator obtained in this procedure and investigate how it relates to the supergravity picture. To separate the $N$ and $M$ D-instanton stacks, we take the 10 (hermitian, traceless) bosonic matrices $X^m$ to be
\begin{align} \label{eq:IKKT_traceless_split_X}
    & X^{(N+M)}_m = \notag \\
    &
    \begin{pmatrix}
        Y^{(N)}_m + \frac{M}{N+M} \frac{\Delta_m}{2\pi\alpha'} \mathbbm{1}_{N} & w_m\\
        w_m^{\dagger} & Y^{(M)}_m - \frac{N}{N+M} \frac{\Delta_m}{2\pi\alpha'} \mathbbm{1}_{M}
    \end{pmatrix} ,
\end{align}
where $Y^{(N)}_m$ and $Y^{(M)}_m$ are traceless and hermitian, and the $w_m$'s are $N\times M$ matrices.

Physically, the diagonal elements represent the positions of the instantons: we pull $N$ of them far from the other $M$ (with center-of-mass separation $\Delta_{\mu}$). Fortunately, IKKT computed the leading deformation to the effective action of the $N$-stack in \cite[§3.4]{Ishibashi:1996xs}:
\begin{equation} \label{eq:IKKT_deformation}
    S \approx S_{\text{IKKT}} + \tilde{c} \Tr [Y,Y]^4 \,,\quad
    \tilde{c} = -\frac{6 M}{(\Delta/2\pi\alpha')^8}\,.
\end{equation}
This result follows from a standard Gaussian integral in which the action is expanded to the quadratic order in the $w_m$'s. This is a good approximation in the limit $\Delta/\alpha' \to \infty$ where the fluctuations of $w_m$'s are asymptotically suppressed.
Interestingly, the same operator can be obtained from dimensional reduction of the operator $\mathcal{O}_8$ in \eqref{Costa-deformation}.
We see that the holographic dictionary, i.e.\ the relation between the coupling of this operator and the constant term in the deformed harmonic \eqref{D-1harm}, takes the same form as in the D3 case, $\tilde{c} \propto c$.

To take this one step further, we consider the D-instanton action coupled to background fields and check how the $\Tr[Y,Y]^4$-operator couples to these. In fact, this is universal for any D$p$-brane as we now show.
We start by expanding the DBI Lagrangian for a stack of D$p$-branes in a general background to the order that is trustworthy
\begin{equation}\label{DBI-expansion}
    \mathcal{L} = \frac{T_p}{e^\phi} \left(\Tr \mathbbm{1} + \frac{(2\pi\alpha')^2}{4} \Tr F^2 - \frac{(2\pi\alpha')^4}{8} \Tr F^4 + \ldots \right)\,,
\end{equation}
where $F$ has $10\mathrm{d}$ indices $F_{MN}$ with $F_{\mu\nu}$ the worldvolume field strength. The transversal directions correspond to the worldvolume scalars, $F_\mu{}^n = D_\mu X^n$ and $F^{mn} = i [X^m, X^n]$, where the eigenvalues of $2\pi\alpha' X^m$ are the positions of the branes.
Note that the field strength terms $F_{\mu\nu}$ are contracted with inverse metrics $g^{\mu\nu}$ along the worldvolume directions but the scalars $X^m$ are contracted with (non-inverted) metrics $g_{mn}$ along the transversal space.
We assume that the background dilaton field $\phi$ and metric $g_{MN}$ only depend on the center-of-mass position of the $N$-stack. In general, they are functionals of the full position matrices $X^m$ and will induce operators like $X^m X^n\nabla_m\nabla_n\phi$ in the DBI action. These operators involve derivatives along the transversal directions, so they are subleading in the large $\Delta/\alpha'$ limit in our setup.

Using the supergravity background $e^{\phi}=g_s H^{(3-p)/4}$, $\sqrt{-g} \sim H^{-(p+1)/4}$ and $g^{\mu\nu} \sim H^{1/2} \sim g_{mn}$, we find the Lagrangian density
\begin{align} \label{eq:DBI_expansion}
    & \sqrt{g} \mathcal{L} = \frac{T_p}{g_s}H^{-1}\Tr \mathbbm{1} + \frac{1}{4 g_\mathrm{YM}^2} \Tr F^2 + \tilde{c} \Tr F^4 + \dots \,, \notag \\
    & g_\mathrm{YM}^2 = (2\pi)^{p-2} g_s (\alpha')^{\frac{p-3}{2}}\,, \quad
    \tilde{c} = - \frac{(2\pi\alpha')^4}{8} \frac{T_p}{g_s} H\,,
\end{align}
where, again, both the worldvolume field strength and scalars are included in the notation but the indices are contracted with the respective flat metrics.
The harmonic in $\tilde{c}$ is evaluated on the stack of $N$ branes and is sourced only by the second stack of $M$ branes, since we are using the open-string description of the $N$-stack.
This reproduces the values for $\tilde{c}$ in \eqref{Costa-deformation} and \eqref{eq:IKKT_deformation} exactly in the D3 and D$(-1)$ cases, respectively.

We interpret the above as follows: the matrix computation gives the coefficient of the leading correction to the IKKT action on the Coulomb branch, which informs us about the form of the harmonic in the dual geometry. In this sense the geometry emerges from the matrix model.

The deformed geometry of D3-branes \eqref{eq:deformed_H} is trustworthy when the string coupling $g_s$ is small but both $g_sN$ and $g_s M$ are large. However, the DBI action \eqref{DBI-expansion} for the $N$-stack captures only the perturbative dynamics $g_s N \to 0$. For the precise match above, only the $c$-term in \eqref{eq:deformed_H} plays a role, so considering $g_s N \to 0$ such that the $N$-stack does not backreact is not a problem.
The D-instanton supergravity background is trustworthy when both the closed string coupling $e^{\phi} = g_s H$ and the curvature $\alpha'R \sim (\Delta/\sqrt{\alpha'})^2(g_s M)^{-1/2}$ are parametrically small.
This means that we are working in the regime
\begin{equation}
    \frac{1}{\sqrt{\alpha'}} g_s^{1/4} M^{1/8} \ll \frac{\Delta}{\alpha'} \ll \frac{1}{\sqrt{\alpha'}} g_s^{1/4} M^{1/4},\quad g_s M \to \infty.
\end{equation}

\section{Matrix RG flows}

We now move on to another proposal for matrix RG flow introduced by Br\'ezin and Zinn-Justin (BZJ) \cite{Brezin_ZJ_1992} where one lowers the rank of the matrices from $N$ to $N-1$ by integrating out a row and a column \footnote{Other proposals for reducing the rank have been put forward, for example averaging over $M\times M$ sub-blocks such that $N\rightarrow N/M$ \cite{Narayan:2002gv}.}. The idea stems from the usual intuition that the number of degrees of freedom decreases as one flows to the IR in QFT. The integrating-out procedure is similar to the Coulomb branch proposal, but we will discuss some important differences. We argue that the $N$-dependence of the dilaton field $e^\phi$ sourced by the $N$-stack follows from the BZJ proposal.

\subsection{Toy model warm-up:}
We illustrate the method with the simple toy-model considered in \cite{Brezin_ZJ_1992}.
Consider the quartic matrix model
\begin{align}
   & Z_N[g,\Lambda] = \int [\dd M_N] e^{-S(M_N;g,\Lambda)}\, ,\nonumber\\
&S = N\left(\frac{1}{2}\Tr M_N^2 + \frac{g}{4}\Tr M_N^4\right) + \Lambda \Tr\mathbbm{1}_N .
\end{align}
The $\Lambda$ term effectively represents the overall normalization of $Z_N$. We will write the $N \times N$ matrix $M_N$ as a $(N-1)\times (N-1)$ matrix in the upper left corner, and integrate out the remaining row and column, perturbatively in $g$. One can then write the remaining integral over $M_{N-1}$, where the coefficients of the constant, Gaussian and quartic terms may be modified. This is the first step of RG.

In the remaining integral over $M_{N-1}$, we will have to rescale $M_{N-1}$, $g$ and $\Lambda$ such that the partition function is of the same form as in the original definition above. This is the second step of RG, namely,
\begin{align}
    Z_N[g,\Lambda] = \int [\dd M_{N-1}'] e^{-S(M_{N-1}', g',\Lambda')} = Z_{N-1}[g',\Lambda']\,.
\end{align}
This is the analog of the Callan--Symanzik (CS) equation, which defines $g(N-1)$ in terms of $g(N)$. In the large $N$ limit, we expect to find smoothly ``running couplings'' as in QFT. Concretely, we have:
\begin{align}
    M_N = \begin{pmatrix}
        M_{N-1} & \mu\\
        \mu^{\dagger} & m
    \end{pmatrix}\, ,
\end{align}
where $\mu$ is a complex $(N-1)\times 1$ vector, and $m$ is a real number. One then obtains:
\begin{align}
    & S(M_N;g,\Lambda) = \notag \\
    &\ N\left(\frac{1}{2}\Tr(M_{N-1}^2)+\frac{g}{4}\Tr(M_{N-1}^4)\right)+ N \Lambda \nonumber  \\
    &\ + N\left(\mu^{\dagger}\mu + \frac{1}{2}m^2\right) +
     g N\mu^{\dagger}\left(M_{N-1}^2 + mM_{N-1}\right)\mu \nonumber \\
    &\ + gN \left(\frac{1}{2} (\mu^{\dagger}\mu)^2
    +  m^2 \mu^{\dagger}\mu
    + \frac{1}{4} m^4 \right)
\end{align}
We recognize the action evaluated on $M_{N-1}$ with the same overall factor $N$ and coupling $g$ as before, a Gaussian term for $\mu$ and $m$, and the interaction terms between $\mu$, $m$ and $M_{N-1}$ which are multiplied by $gN$.

Let us first look at the classical dimension of the coupling. For that we require that the classical part of the action remains invariant. So we equate
\begin{align}
 &N\left(\frac{1}{2}\Tr(M_{N-1}^2)+\frac{g}{4}\Tr(M_{N-1}^4)\right) = \notag \\
 & (N-1)\left(\frac{1}{2}\Tr(M^{\prime 2}_{N-1})+\frac{g'}{4}\Tr(M^{\prime 4}_{N-1})\right)\,.
\end{align}
From comparing the Gaussian terms and quartic terms we find respectively:
\begin{align}
M^{\prime 2}_{N-1} &= \frac{N}{N-1}M_{N-1}^2\,, \quad g'=\frac{N-1}{N}g\,.    \label{eq:rescaling_M_to_Mprime}
\end{align}
To get the leading order correction to the ``classical scaling'' of $g$, we integrate out $m$, $\mu$ and $\mu^{\dagger}$ and eventually find: \begin{align}
    g' &= g \frac{N-1}{N}\left(1-\frac{6g}{N}+\mathcal{O}(g^2)\right)\,.
\end{align}
By tracking the overall normalization of $Z$ one can also compute a running of the ``vacuum energy'' $\Lambda'$ at every order in $g$.

\subsection{Matrix RG for IKKT:}
If one ignores the $g$-dependence of the path integral normalization factor $\mathcal{N}$, then the coupling $g$ of the IKKT action can be redefined away by rescaling $X$ and $\psi$ as $X = \sqrt{g}\tilde{X}\,,\quad \psi = g^{3/4}\tilde{\psi}$. Furthermore, the absence of a Gaussian term seems an obstacle for applying the BZJ method. Below we address both the $(g,N)$-dependence of $\mathcal{N}$ and the lack of a Gaussian term. We first start with the $(g,N)$-dependence of $\mathcal{N}$, which is fixed by the context in which one uses the IKKT model. For our case this context is D-instanton physics.

\subsubsection{Normalising the partition function}

Crucially, the context of D-instanton physics, for which the IKKT action is being used can fix the $(g, N)$-dependence of $\mathcal{N}$ uniquely \cite{Vanhove:1999qw,Sen:2021jbr}.
We recall that the D-instanton contributes to the $\mathcal{R}^4$-term in the IIB action. First, the contributions are associated with a weight $e^{-S}$, where $S$ is the classical action of D-instantons
\begin{equation}\label{D-instanton-action}
S = \frac{2\pi N}{e^{\phi}} + i2\pi NC_0\,,
\end{equation}
with $C_0$ the RR axion of IIB. To capture this in IKKT one can simply add a ``tree-level vacuum energy'' term to the IKKT action (which also follows from expanding the DBI action; see \eqref{eq:DBI_expansion}):
\begin{equation}
\label{eq:IKKT_with_trace_1}
S=2\pi \left(e^{-\phi} +i C_0\right)\Tr\mathbbm{1}+S_{\text{IKKT}}   + \text{higher order}\,,
\end{equation}
where the coupling in IKKT action is given by $g^2 = (2\pi)^{-3} e^{\phi} (\alpha')^{-2}$. Henceforth, for notational simplicity, we denote $e^{\phi}=g_s$ in this section, not to be confused with the asymptotic value of the dilaton from previous sections.

The tension term fixes the physical meaning of the IKKT coupling. Next, a factor of $(N/g_s)^{1/2}$ appears in the multi-instanton contribution to the four graviton scattering \cite{Vanhove:1999qw,Sen:2021jbr}. This fixes the normalization of the $U(1)$ part of the matrix model, and the result is that the D-instanton partition function carries a factor of $(g_s/N)^{7/2}$ \cite{Vanhove:1999qw} (accompanied with a factor of $(N/g_s)^4$ from the external gravitons to reproduce the $(N/g_s)^{1/2}$ factor).

A second point of confusion in the literature is the gauge volume dependence in the normalization $\mathcal{N}$. The ungauged IKKT model has an extra $N$-dependent factor stemming from integrating over gauge orbits. However, a match with string theory computations requires one to use the gauged model and remove this term from the partition function. After fixing the normalization, the final result of the partition function is (modified from \cite{Vanhove:1999qw})
\begin{equation}\label{TheproperZ}
Z(N, g_s) = e^{-\frac{2\pi N}{g_s} - 2\pi i N C_0}\, \sigma_{-2}(N) \, \left(\frac{g_s}{N}\right)^{7/2}\,,
\end{equation}
with the divisor function $\sigma_{-2}(N)$,
\begin{equation}
\sigma_{-2}(N) = \sum_{d|N} \frac{1}{d^2}\,,
\end{equation}
computed from the matrix integral in \cite{Moore:1998et}.

\subsubsection{Coarse-graining the partition function}

The perturbative approach to the BZJ flow does not work for the IKKT model because of the absence of a Gaussian term. One way out is to use the analog to the BZJ flow on the Coulomb branch $SU(N)\to SU(N-1)\times U(1)$. We will not follow this line because of some difficulties, but simply report an interesting observation on the flow of the undeformed IKKT partition function \eqref{TheproperZ}, where the ``Coulomb branch position'' is integrated over.

In order to take derivatives and apply matrix RG flows, we need to coarse-grain $\sigma_{-2}(N)$. This is akin to what has been discussed in \cite{Liu:2025ikq,Kudler-Flam:2026nzz}. In the appendix we present this computation of which the outcome is that the derivative of $\sigma_{-2}(N)$  seems to vanish at leading order at large $N$. With this in mind we apply a Callan--Symanzik (CS) equation to the partition function:
\begin{align}\label{flow-part}
    N\frac{\dd}{\dd N} Z_{\text{IKKT}}(N,g_s(N), C_0(N)) = 0\,.
\end{align}

Compared to the BZJ flow of the toy model, the flow \eqref{flow-part} is not based on the effective action. This raises the concern of the validity of the flow derived from \eqref{flow-part} when used in expectation values. We expect that, based on the Coulomb branch computation, an infinite tower of ``irrelevant operators'' will be generated, but an explicit computation is beyond our ability. Therefore, we do not claim the flow generated from \eqref{flow-part} to be consistent with the CS equation for expectation values.
However, unlike the toy model, we have here the exact partition function at each $N$. This allows us to study the BZJ flow of the $\Tr \mathbbm{1}$ term precisely.
Note that the running of $C_0$, which fixes the complex phase of $Z$ in \eqref{TheproperZ}, is trivial:
\begin{equation}
C_0(N)\sim N^{-1}\,,
\end{equation}
Note that the supergravity profile of the RR axion in the near-horizon region behaves as $C_0(r) \sim r^8N^{-1}$, which has the same $N$ scaling as the above BZJ flow result. For understanding the $N$ flow of $g_s$ we compute the renormalized vacuum energy by expanding the CS equation:
\begin{equation}
\left(\frac{2\pi}{g_s^2} +\frac{7}{2 g_s N}\right) N \frac{dg_s}{dN} = \frac{2\pi}{g_s}-\frac{1}{\sigma_{-2}(N)}\frac{d \sigma_{-2}}{dN} +\frac{7}{2N}\,.
\end{equation}
Hence, in the large $N$ limit, we have
\begin{equation}
N \frac{dg_s}{dN} \sim g_s - g_s^2\frac{1}{2\pi\sigma_{-2}(N)}\frac{d \sigma_{-2}}{dN}\,.
\end{equation}
Since  $\mathrm{d}\sigma_{-2}(N)/\mathrm{d}N = 0$ after the coarse-graining, we find in the large $N$ limit:
\begin{align}
g_s(N) \approx g_s(N_0) \frac{N}{N_0}\,.
\end{align}
As for $C_0$, one can see from the supergravity side that the dilaton profile in the near-horizon limit of a stack of $N$ $D$-instantons behaves like $e^{\phi} \sim N/r^8$, so the power of $N$ is the same as the $g_s$ flow found above.

The main point is that the naive ``classical'' scaling of $g_s$ with $N$ that one can obtain by enforcing \eqref{eq:IKKT_with_trace_1} to be $N$-independent is not corrected at the ``quantum'' level (i.e.\ when taking into account the full partition function \eqref{TheproperZ}). However, it remains to be seen whether this continues to hold when accounting for the operators generated along the flow, tracking the full effective action instead of the partition function. Because of maximal supersymmetry, we suspect that the analog of the BPS bound which protects D-brane tensions applies to this $N$-flow.

\section{Discussion}

As a way to explore the possibility of 0-dimensional holography, we proposed in this letter two methods for reducing the rank of the matrix theory. The hope is that these ``matrix RG flows'' would be dual to radial flow of fields on the gravity side.

To summarize, we first proposed a method to uncover an emergent radial direction by going on the Coulomb branch. This can be implemented in IKKT by expanding around a fixed background corresponding to two separated stacks of D-instantons at a fixed distance $\Delta$. The leading correction to IKKT as $\Delta \to \infty$ matches the constant piece in the harmonic of the two-centered D-instanton supergravity background. Moreover, we find the same constant piece from the DBI action on the near-horizon D-instanton background. In this way, we extract the profile of the dilaton in the D-instanton geometry from a matrix theory computation.

A closely related approach we used is inspired by \cite{Brezin_ZJ_1992} and computes the $N$-dependence of the IKKT coupling constant $g_{\text{IKKT}}(N)$ and the IKKT theta-angle $\theta_{\text{IKKT}}$. The $N$-scaling of these matrix couplings exactly match the $N$-scaling of the (near-horizon) dilaton profile and the RR axion $C_0$ in the bulk. This method relies on completely integrating out rows and columns and so differs from the previous method by further integrating over all Coulomb branch positions and imposing a Callan--Symanzik-like equation on the partition function to induce a running of the couplings with $N$.

Whereas these procedures derive the radial and $N$-dependence of the supergravity fields sourced by a stack of D-instantons, the emergence of the angles of the $S^9$ occurs through the $SO(10)$ labels of IKKT multiplets, as explained in \cite{Ciceri:2025maa}. Together, these results give a conceptual meaning to IKKT holography and to a potential start of more in-depth checks on both sides of the duality by deforming the background with sources. Aside from deformations with operators carrying a non-trivial $SO(10)$-representation, one can look at deformations with the $B_2$- (or $C_2$)-field which makes the instantons puff into spherical strings \cite{Komatsu:2024bop, Komatsu:2024ydh, Hartnoll:2024csr}. Our procedure then suggests computing the dependence of those couplings on the distance between two stacks.

A related exciting challenge is to reproduce the constant dilaton-profile in the near-horizon of a $D(-1)/D7$ bound state solution \cite{Reymond:2024mwe} which indicates a fixed point in the BZJ flow \cite{Aguilar-Gutierrez:2022kvk}. The would-be dual conformal matrix theory is derived from adding $D7$ interactions to the IKKT matrix theory (``Higgs branch'') and is explicitly derived in \cite{Billo:2021xzh}. We hope to come back to this in the future.

\begin{acknowledgments}
\emph{ We are very grateful to Jo\~ao Melo for collaboration at early stages and Nikolay Bobev for critical remarks throughout the project. We furthermore acknowledge useful discussions and correspondence with Guillermo Mera {\'A}lvarez, Jo\~ao Penedones and Xiang Zhao. The work of T.V.R.\ and J.K.\ is supported by the KU~Leuven C1 grant
ZKE7799C16/25/010. J.K.\ is further supported by the Research Foundation - Flanders (FWO) doctoral fellowship 1171825N. S.R.\ is supported by the FWO doctoral fellowship 11PAA24N, and also
acknowledges support from the FWO Odysseus grant G0F9516N and the KU Leuven iBOF-21-084 grant.}
\end{acknowledgments}

\appendix

\section{Coarse graining the divisor function}
One coarse graining strategy is to switch to the grand canonical ensemble, and extract the large $N$ scaling from a small chemical potential asymptotic expansion. The grand canonical partition function is defined as
\begin{equation} \label{eq:NM_Laplace}
    \mathcal{Z}_\mathrm{NM}(\mu)
    = \sum_{N = 1}^\infty \sigma_{-2}(N) e^{-\mu N}
    = \sum_{m = 1}^\infty \frac{m^{-2}}{e^{m \mu} - 1}\,.
\end{equation}
The sum is convergent for $\Re \mu >0$. The divergence behavior as $\mu \to 0$ captures the coarse-grained large $N$ scaling of $\sigma_{-2}(N)$.

The inverse Laplace transformation of \eqref{eq:NM_Laplace} formally recovers the divisor function as a distribution
\begin{equation} \label{eq:NM_inverse_Laplace}
    \frac{1}{2\pi i} \int_I \mathrm{d}\mu\, e^{\mu N} \mathcal{Z}_\mathrm{NM}(\mu)
    = \sum_{n = 1}^\infty \delta(N-n) \sigma_{-2}(n)\,,
\end{equation}
for a contour $I$ parallel to the imaginary axis passing to the right of the origin. On the other hand, the coarse graining procedure replaces $\mathcal{Z}_\mathrm{NM}(\mu)$ with its asymptotic expansion at $\mu \to 0$ in \eqref{eq:NM_inverse_Laplace}.

Because the $\mu \to 0$ limit is singular, one is not allowed to do it for each summand in \eqref{eq:NM_Laplace}. We will instead perform a Mellin transformation
\begin{equation}
    \mathcal{M}[\mathcal{Z}_\mathrm{NM}](s) = \int_0^\infty \mathrm{d}\mu\, \mu^{s-1} \mathcal{Z}_\mathrm{NM}(\mu)
    = \Gamma(s) \zeta(s) \zeta(s+2)\,,
\end{equation}
and transform it back after an analytic continuation in $s$
\begin{equation}
    \mathcal{Z}_\mathrm{NM}(\mu) = \frac{1}{2\pi i} \int_I \mathrm{d}s\, \mu^{-s} \mathcal{M}[\mathcal{Z}_\mathrm{NM}](s)\,,
\end{equation}
where the contour is closed to the left and picking up the poles. The leading singularity comes from the simple pole at $s=1$ whose residue is $\zeta(3)$
\begin{equation}
    \mathcal{Z}_\mathrm{NM}(\mu) \stackrel{\mu\to0}{\sim} \frac{\zeta(3)}{\mu} + \dots\,.
\end{equation}
Other possible poles due to the $\Gamma(s)$ locates at non-positive integers $s=0,-1,-2,\dots$, but may be killed by the zeros of the $\zeta$-function. Including the contributions from all the poles, one obtains the asymptotic expansion
\begin{align}\label{asymp}
    \mathcal{Z}_\mathrm{NM}(\mu) \sim
    \frac{\zeta(3)}{\mu} - \frac{\pi^2}{12} - \frac{1}{12} \mu \log \frac{\mu}{A^{12}} \notag\\
    + \sum_{k=3}^\infty (-1)^k \frac{\zeta(-k) \zeta(2-k)}{k!} \mu^k\,.
\end{align}

We now evaluate the inverse Laplace transformation \eqref{eq:NM_inverse_Laplace} with the asymptotic expansion \eqref{asymp} using a saddle point approximation at large $N$. From the saddle point equation
\begin{equation}
    \frac{\partial}{\partial \mu} \mathcal \log Z_{\mathrm{NM}}(\mu) + N = 0,
\end{equation}
one finds a saddle point located at
\begin{equation}
    \mu_\ast = N^{-1} + \frac{\pi^2}{12 \zeta(3)} N^{-2} + \mathcal{O}(N^{-3} \log N)\, .
\end{equation}
Now, the contribution from this saddle point to the inverse Laplace transform is
\begin{equation} \label{eq:NM_iL_saddle_approx}
    \tilde{Z}_\mathrm{NM}(N) = \frac{e^{\mu_\ast N} \mathcal{Z}_\mathrm{NM}(\mu_\ast)}{2\pi i} \int_I \mathrm{d}\mu\, \exp\biggl[\sum_{k \geq 2} c_k\, \delta\mu^k\biggr]\,,
\end{equation}
with
\begin{equation*}
    c_k = (-1)^k\frac{N^k}{k}\left( 1-\frac{\pi^2}{12\zeta(3)}\frac{k}{N} + O(N^{-2})\right).
\end{equation*}

At the leading order of large $N$, we have
\begin{equation}
    e^{\mu_\ast N} \mathcal{Z}_\mathrm{NM}(\mu_\ast) \sim e \zeta(3) N + \dots\,,
\end{equation}
and
\begin{equation}
    \sum_{k \geq 2} c_k\, \delta\mu^k
    \sim N \delta\mu - \log(1 + N \delta\mu) + \dots\,.
\end{equation}
Performing the integral in \eqref{eq:NM_iL_saddle_approx} by picking up the pole at $\delta\mu = -1/N$ (corresponding to $\mu = 0$) gives
\begin{equation}
    \tilde{Z}_\mathrm{NM}(N) \sim \zeta(3) + \dots\,.
\end{equation}
Therefore, the large $N$ coarse graining of the divisor function $\sigma_{-2}(N)$ yields a constant function.

\bibliographystyle{utphys-modified}
\bibliography{refs}

\providecommand{\href}[2]{#2}\begingroup\raggedright\begin{thebibliography}{10}

\bibitem{Ishibashi:1996xs}
N.~Ishibashi, H.~Kawai, Y.~Kitazawa, and A.~Tsuchiya, ``{A Large N reduced
  model as superstring},''
  \href{http://dx.doi.org/10.1016/S0550-3213(97)00290-3}{{\em Nucl. Phys. B}
  {\bfseries 498} (1997) 467--491},
  \href{http://arxiv.org/abs/hep-th/9612115}{{\ttfamily arXiv:hep-th/9612115}}.

\bibitem{Note1}
We adopt the notation $[X, X]^2 = [X_m, X_n] [X^m, X^n]$. The Weyl spinors
  $\psi $ have $16$ complex components and $\psi \Gamma _m [X^m, \psi ] = \psi
  _\alpha (\protect \mathcal {C} \Gamma ^m)^{\alpha \beta } [X^m,\psi _\beta
  ]$. The convention for the fermion term is discussed in \cite
  {Hartnoll:2024csr,Komatsu:2024bop}.

\bibitem{Gibbons:1995vg}
G.~W. Gibbons, M.~B. Green, and M.~J. Perry, ``{Instantons and seven-branes in
  type IIB superstring theory},''
  \href{http://dx.doi.org/10.1016/0370-2693(95)01565-5}{{\em Phys. Lett. B}
  {\bfseries 370} (1996) 37--44},
  \href{http://arxiv.org/abs/hep-th/9511080}{{\ttfamily arXiv:hep-th/9511080}}.

\bibitem{Nishimura:2020blu}
J.~Nishimura, ``{New perspectives on the emergence of (3+1)D expanding
  space-time in the Lorentzian type IIB matrix model},''
  \href{http://dx.doi.org/10.22323/1.376.0178}{{\em PoS} {\bfseries CORFU2019}
  (2020) 178}, \href{http://arxiv.org/abs/2006.00768}{{\ttfamily
  arXiv:2006.00768 [hep-lat]}}.

\bibitem{Brahma:2022dsd}
S.~Brahma, R.~Brandenberger, and S.~Laliberte, ``{Emergent metric space-time
  from matrix theory},'' \href{http://dx.doi.org/10.1007/JHEP09(2022)031}{{\em
  JHEP} {\bfseries 09} (2022) 031},
  \href{http://arxiv.org/abs/2206.12468}{{\ttfamily arXiv:2206.12468
  [hep-th]}}.

\bibitem{Steinacker:2024unq}
H.~C. Steinacker, \href{http://dx.doi.org/10.1017/9781009440776}{{\em {Quantum
  Geometry, Matrix Theory, and Gravity}}}.
\newblock Cambridge University Press, 4, 2024.

\bibitem{Steinacker:2026jzp}
H.~C. Steinacker, ``{Quantum spacetime and quantum fluctuations in the IKKT
  model at weak coupling},'' \href{http://arxiv.org/abs/2605.13294}{{\ttfamily
  arXiv:2605.13294 [hep-th]}}.

\bibitem{Ciceri:2025maa}
F.~Ciceri and H.~Samtleben, ``{Holography for the
  Ishibashi-Kawai-Kitazawa-Tsuchiya Matrix Model},''
  \href{http://dx.doi.org/10.1103/fb8g-b8fd}{{\em Phys. Rev. Lett.} {\bfseries
  135} no.~6, (2025) 061601}, \href{http://arxiv.org/abs/2503.08771}{{\ttfamily
  arXiv:2503.08771 [hep-th]}}.

\bibitem{Ciceri:2025wpb}
F.~Ciceri and H.~Samtleben, ``{Supergravity dual for
  Ishibashi-Kawai-Kitazawa-Tsuchiya holography},''
  \href{http://dx.doi.org/10.1103/gmhl-mmg5}{{\em Phys. Rev. D} {\bfseries 113}
  no.~4, (2026) 046001}, \href{http://arxiv.org/abs/2511.23111}{{\ttfamily
  arXiv:2511.23111 [hep-th]}}.

\bibitem{Ciceri:2026hxr}
F.~Ciceri and H.~Samtleben, ``{The matrix edge of holography},''
  \href{http://arxiv.org/abs/2604.00355}{{\ttfamily arXiv:2604.00355
  [hep-th]}}.

\bibitem{Brezin_ZJ_1992}
E.~Brézin and J.~Zinn-Justin, ``Renormalization group approach to matrix
  models,'' \href{http://dx.doi.org/10.1016/0370-2693(92)91953-7}{{\em Physics
  Letters B} {\bfseries 288} no.~1–2, (Aug., 1992) 54–58}.
  \url{http://dx.doi.org/10.1016/0370-2693(92)91953-7}.

\bibitem{Higuchi:1993pu}
S.~Higuchi, C.~Itoi, and N.~Sakai, ``{Renormalization group approach to matrix
  models and vector models},''
  \href{http://dx.doi.org/10.1143/PTPS.114.53}{{\em Prog. Theor. Phys. Suppl.}
  {\bfseries 114} (1993) 53--71},
  \href{http://arxiv.org/abs/hep-th/9307154}{{\ttfamily arXiv:hep-th/9307154}}.

\bibitem{Zinn-Justin:2014wva}
J.~Zinn-Justin, ``{Random vector and matrix and vector theories: a
  renormalization group approach},''
  \href{http://dx.doi.org/10.1007/s10955-014-1103-y}{{\em J. Statist. Phys.}
  {\bfseries 157} (2014) 990--1016},
  \href{http://arxiv.org/abs/1410.1635}{{\ttfamily arXiv:1410.1635 [math-ph]}}.

\bibitem{Ydri:2020efr}
B.~Ydri and R.~Ahmim, ``{Wilsonian renormalization group for a multitrace
  matrix model},'' \href{http://dx.doi.org/10.1142/S0217751X22501652}{{\em Int.
  J. Mod. Phys. A} {\bfseries 37} no.~27, (2022) 2250165},
  \href{http://arxiv.org/abs/2008.09564}{{\ttfamily arXiv:2008.09564
  [hep-th]}}.

\bibitem{Lahoche:2019ocf}
V.~Lahoche and D.~Ousmane~Samary, ``{Revisited functional renormalization group
  approach for random matrices in the large-$N$ limit},''
  \href{http://dx.doi.org/10.1103/PhysRevD.101.106015}{{\em Phys. Rev. D}
  {\bfseries 101} no.~10, (2020) 106015},
  \href{http://arxiv.org/abs/1909.03327}{{\ttfamily arXiv:1909.03327
  [hep-th]}}.

\bibitem{Narayan:2002gv}
K.~Narayan, ``{Blocking up D branes: Matrix renormalization?},''
  \href{http://arxiv.org/abs/hep-th/0211110}{{\ttfamily arXiv:hep-th/0211110}}.

\bibitem{Itzhaki:1998dd}
N.~Itzhaki, J.~M. Maldacena, J.~Sonnenschein, and S.~Yankielowicz,
  ``{Supergravity and the large N limit of theories with sixteen
  supercharges},'' \href{http://dx.doi.org/10.1103/PhysRevD.58.046004}{{\em
  Phys. Rev. D} {\bfseries 58} (1998) 046004},
  \href{http://arxiv.org/abs/hep-th/9802042}{{\ttfamily arXiv:hep-th/9802042}}.

\bibitem{Boonstra:1998mp}
H.~J. Boonstra, K.~Skenderis, and P.~K. Townsend, ``{The domain wall / QFT
  correspondence},''
  \href{http://dx.doi.org/10.1088/1126-6708/1999/01/003}{{\em JHEP} {\bfseries
  01} (1999) 003}, \href{http://arxiv.org/abs/hep-th/9807137}{{\ttfamily
  arXiv:hep-th/9807137}}.

\bibitem{Peet:1998wn}
A.~W. Peet and J.~Polchinski, ``{UV-IR relations in AdS dynamics},''
  \href{http://dx.doi.org/10.1103/PhysRevD.59.065011}{{\em Phys. Rev. D}
  {\bfseries 59} (1999) 065011},
  \href{http://arxiv.org/abs/hep-th/9809022}{{\ttfamily arXiv:hep-th/9809022}}.

\bibitem{Kanitscheider:2008kd}
I.~Kanitscheider, K.~Skenderis, and M.~Taylor, ``{Precision holography for
  non-conformal branes},''
  \href{http://dx.doi.org/10.1088/1126-6708/2008/09/094}{{\em JHEP} {\bfseries
  09} (2008) 094}, \href{http://arxiv.org/abs/0807.3324}{{\ttfamily
  arXiv:0807.3324 [hep-th]}}.

\bibitem{Bobev:2025idz}
N.~Bobev, G.~Mera~{\'A}lvarez, and H.~Paul, ``{Correlation functions for
  non-conformal Dp-brane holography},''
  \href{http://dx.doi.org/10.1007/JHEP07(2025)137}{{\em JHEP} {\bfseries 07}
  (2025) 137}, \href{http://arxiv.org/abs/2503.18770}{{\ttfamily
  arXiv:2503.18770 [hep-th]}}.

\bibitem{Biggs:2023sqw}
A.~Biggs and J.~Maldacena, ``{Scaling similarities and quasinormal modes of D0
  black hole solutions},''
  \href{http://dx.doi.org/10.1007/JHEP11(2023)155}{{\em JHEP} {\bfseries 11}
  (2023) 155}, \href{http://arxiv.org/abs/2303.09974}{{\ttfamily
  arXiv:2303.09974 [hep-th]}}.

\bibitem{Costa:2000gk}
M.~S. Costa, ``{A Test of the AdS / CFT duality on the Coulomb branch},''
  \href{http://dx.doi.org/10.1016/S0370-2693(00)00484-6}{{\em Phys. Lett. B}
  {\bfseries 482} (2000) 287--292},
  \href{http://arxiv.org/abs/hep-th/0003289}{{\ttfamily arXiv:hep-th/0003289}}.
  [Erratum: Phys.Lett.B 489, 439--439 (2000)].

\bibitem{Note2}
$\Tr F^4 = \protect \frac {1}{3} \Tr (F_{mn} F^{np} F_{pq} F^{qm}) + \protect
  \frac {2}{3} \Tr (F_{mn} F_{pq} F^{mq} F^{pn}) - \protect \frac {1}{6} \Tr
  (F_{mn} F^{mn} F_{pq} F^{pq}) - \protect \frac {1}{12} \Tr (F_{mn} F_{pq}
  F^{mn} F^{pq})$.

\bibitem{Note3}
Other proposals for reducing the rank have been put forward, for example
  averaging over $M\times M$ sub-blocks such that $N\rightarrow N/M$ \cite
  {Narayan:2002gv}.

\bibitem{Vanhove:1999qw}
P.~Vanhove, ``{D instantons and matrix models},''
  \href{http://dx.doi.org/10.1088/0264-9381/16/10/308}{{\em Class. Quant.
  Grav.} {\bfseries 16} (1999) 3147--3164},
  \href{http://arxiv.org/abs/hep-th/9903050}{{\ttfamily arXiv:hep-th/9903050}}.

\bibitem{Sen:2021jbr}
A.~Sen, ``{Muti-instanton amplitudes in type IIB string theory},''
  \href{http://dx.doi.org/10.1007/JHEP12(2021)065}{{\em JHEP} {\bfseries 12}
  (2021) 065}, \href{http://arxiv.org/abs/2104.15110}{{\ttfamily
  arXiv:2104.15110 [hep-th]}}.

\bibitem{Moore:1998et}
G.~W. Moore, N.~Nekrasov, and S.~Shatashvili, ``{D-particle bound states and
  generalized instantons},''
  \href{http://dx.doi.org/10.1007/s002200050016}{{\em Commun. Math. Phys.}
  {\bfseries 209} (2000) 77--95},
  \href{http://arxiv.org/abs/hep-th/9803265}{{\ttfamily arXiv:hep-th/9803265}}.

\bibitem{Liu:2025ikq}
H.~Liu, ``{''Filtering'' CFTs at large N: Euclidean Wormholes, Closed
  Universes, and Black Hole Interiors},''
  \href{http://arxiv.org/abs/2512.13807}{{\ttfamily arXiv:2512.13807
  [hep-th]}}.

\bibitem{Kudler-Flam:2026nzz}
J.~Kudler-Flam and E.~Witten, ``{Wormholes and Averaging over N},''
  \href{http://arxiv.org/abs/2605.15180}{{\ttfamily arXiv:2605.15180
  [hep-th]}}.

\bibitem{Komatsu:2024bop}
S.~Komatsu, A.~Martina, J.~a. Penedones, A.~Vuignier, and X.~Zhao, ``{Einstein
  gravity from a matrix integral -- Part I},''
  \href{http://arxiv.org/abs/2410.18173}{{\ttfamily arXiv:2410.18173
  [hep-th]}}.

\bibitem{Komatsu:2024ydh}
S.~Komatsu, A.~Martina, J.~Penedones, A.~Vuignier, and X.~Zhao, ``{Einstein
  gravity from a matrix integral -- Part II},''
  \href{http://arxiv.org/abs/2411.18678}{{\ttfamily arXiv:2411.18678
  [hep-th]}}.

\bibitem{Hartnoll:2024csr}
S.~A. Hartnoll and J.~Liu, ``{The Polarised IKKT Matrix Model},''
  \href{http://arxiv.org/abs/2409.18706}{{\ttfamily arXiv:2409.18706
  [hep-th]}}.

\bibitem{Reymond:2024mwe}
S.~Reymond, M.~Trigiante, and T.~Van~Riet, ``{Supergravity solutions for Dp-D(6
  {\ensuremath{-}} p) bound states: from p = 7 to p = {\ensuremath{-}}1},''
  \href{http://dx.doi.org/10.1007/JHEP04(2025)037}{{\em JHEP} {\bfseries 04}
  (2025) 037}, \href{http://arxiv.org/abs/2410.02616}{{\ttfamily
  arXiv:2410.02616 [hep-th]}}.

\bibitem{Aguilar-Gutierrez:2022kvk}
S.~E. Aguilar-Gutierrez, K.~Parmentier, and T.~Van~Riet, ``{Towards an
  {\textquotedblleft}AdS$_{1}$/CFT$_{0}${\textquotedblright} correspondence
  from the D({\ensuremath{-}}1)/D7 system?},''
  \href{http://dx.doi.org/10.1007/JHEP09(2022)249}{{\em JHEP} {\bfseries 09}
  (2022) 249}, \href{http://arxiv.org/abs/2207.13692}{{\ttfamily
  arXiv:2207.13692 [hep-th]}}.

\bibitem{Billo:2021xzh}
M.~Bill{\`o}, M.~Frau, F.~Fucito, L.~Gallot, A.~Lerda, and J.~F. Morales, ``{On
  the D({\textendash}1)/D7-brane systems},''
  \href{http://dx.doi.org/10.1007/JHEP04(2021)096}{{\em JHEP} {\bfseries 04}
  (2021) 096}, \href{http://arxiv.org/abs/2101.01732}{{\ttfamily
  arXiv:2101.01732 [hep-th]}}.

\end{thebibliography}\endgroup
\end{document}